# Strong optical force acting on a dipolar particle over a multilayer substrate

Shubo Wang and C. T. Chan[*]

*Department of Physics and Institute for Advanced Study,
The Hong Kong University of Science and Technology, Hong Kong, China*

**Abstract**

Optical forces acting on nano-sized particles are typically too small to be useful for particle manipulation. We theoretically and numerically demonstrate a mechanism that can significantly enhance the optical force acting on a small particle through a special type of resonant particle-substrate coupling. The resonance arises from the singular behavior of the particle's effective polarizablity in the presence of a metal-dielectric-metal multilayer substrate. We show that this phenomenon is closely related to the existence of a flat-band plasmon mode supported by the multilayer substrate.

## 1. Introduction

Using optical force to manipulate small particles becomes feasible after the invention of lasers because the optical force is usually not strong enough unless high intensity fields are provided by focusing a laser beam [1, 2]. Such applications are typically limited to micro-sized particles that have reasonably large electric dipole polarizablity. Realizing strong optical forces for manipulating deep sub-wavelength nano-sized (such as tens of nanometer) particles would be highly desirable, but there are many challenges. For structures with a particular shape such as parallel-plate cavity [3, 4], nanowire pairs [5, 6], coupled waveguides [7, 8] and metamaterial slabs [9], it is possible to achieve strong optical forces through cavity resonance or evanescent-field manipulation. However, it is not easy to induce a strong force on a small simple-shaped particle like a nano-sphere [10-12]. It has been shown that when a small particle is put near to a dielectric or metallic substrate, the near-field evanescent wave can induce a gradient force on the particle whose magnitude is much larger than the normal photon pressure [13-20]. In this paper we go one step further and show that if a dipolar particle is put near to a substrate with a metal-dielectric-metal sandwich structure, it is possible to enhance the optical force acting on the particle by orders of magnitude compared to usual evanescent wave induced forces. The enhancement is attributed to a particle-substrate resonance that happens under the condition that the surface plasmon modes supported by the slab has a very flat dispersion, in which case the effective dipole polarizability of the particle shows a divergence feature.

We consider a spherical particle of radius $r$ located at a distance $d$ above a substrate with a three-layer metal-dielectric-metal sandwich structure, as shown in Fig.1. The thicknesses of the three layers are $t_1$, $t_2$, and $t_3$, respectively, and we take $t_3 \to \infty$ so the bottom layer is essentially semi-infinite along $z$ direction. The whole sandwich structure is infinitely large on the $xy$ plane. The relative permittivities of the layer 1 and 3 of the substrate are both given by the Drude model: $\varepsilon_1 = \varepsilon_3 = 1 - \omega_p^2/(\omega^2 + i\omega\gamma)$, where $\omega_p$ is the bulk plasma frequency and $\gamma$ denotes damping frequency. The material of the layer 2 is dielectric with $\varepsilon_2 = 12.96$. We assume all the materials are non-magnetic. The system is excited by an electromagnetic plane wave of $\mathbf{E}_{\mathrm{inc}} = \hat{x} E_0 \mathrm{Exp}[-ik_0 z - i\omega t]$ with an intensity of $1\,\mathrm{mW}/(\mu\mathrm{m})^2$. We will show that under certain conditions the system can support a peculiar particle-substrate resonance that dramatically magnifies the optical force.

The paper is organized as follows. In Sec. 2, we introduce in detail the analytical Green's function method for solving the considered problem. In Sec. 3, we show the results of the force enhancement for two kinds of dipolar particle with either positive or negative polarizability. The conclusions are drawn in Sec. 4.

## 2. Analytical Green's function method

Let us first consider address the problem analytically using the Green's function method. Under the long wavelength condition the particle can be treated as an electric dipole with polarizabiltiy $\alpha^e = i6\pi\varepsilon_0 a_1/k_0^3$, where $a_1$ is the Mie scattering coefficient and $k_0$ is the wavenumber in vacuum [21, 22]. The induced electric dipole moment is

$$\mathbf{p} = \alpha^e \mathbf{E} = \alpha^e \left[ \mathbf{E}^{\mathrm{inc}} \left(1 + R^{\mathrm{TE}} \mathrm{e}^{i2k_0 d}\right) + \mu_0 \omega^2 \bar{\mathbf{G}} \cdot \mathbf{p} \right], \tag{1}$$

where $\mathbf{E}$ is the local field at the position of the dipole $(0,0,d)$; $R^{\mathrm{TE}}$ is the reflection coefficient of the substrate under TE polarization; $\bar{\mathbf{G}}$ is the Green's function accounting for the effect of the substrate. Solve the above equation and we obtain

$$\mathbf{p} = \alpha^{e}_{\text{eff}} \mathbf{E}_{\text{inc}} = \frac{\alpha^{e}\left(1 + R^{\text{TE}} e^{i2k_0 d}\right)}{1 - \mu_0 \omega^2 \alpha^{e} G_{xx}} \mathbf{E}^{\text{inc}} , \tag{2}$$

where $G_{xx}$ is the element of the dyadic Green's function $\overline{\mathbf{G}}$ and we define $\alpha^{e}_{\text{eff}}$ as the effective polarizability (the polarizablity of the particle taking into account of the effect of substrate). We note that the denominator of the above expression is determined by the bare polarizability $\alpha^{e}$ as well as the Green's function of the substrate. If we carefully design the system so that the condition $|1 - \mu_0 \omega^2 \alpha^{e} G_{xx}| = 0$ is fulfilled, then the effective polarizability will diverge, so is the induced dipole moment. Such a pole can be interpreted as a collective particle-substrate resonance mode that only exists when a dipole particle and a substrate are put together. In this case the optical force acting on the particle will be strongly enhanced. We note that the condition $|1 - \mu_0 \omega^2 \alpha^{e} G_{xx}| = 0$ cannot be fulfilled for a normal configuration of particle-over-substrate, say a small particle sitting on a semi-infinite metal substrate, due to the small value of $\alpha^{e}$ and $G_{xx}$. It has been shown that the condition can be reached if the substrate is made of negative-refractive-index materials with $\varepsilon \to -1, \mu \to -1$[23, 24]. However, such material cannot be found in nature and can only be realized by using artificial metamaterials [24-26]. Here we will show that the sandwich structure shown in Fig.1 provides an alternate way to obtain a large value of $G_{xx}$.

The physical meaning of $G_{xx}$ is the interaction between the particle and the substrate induced by multiple reflections. $G_{xx}$ can have a large value if the scattered channels (Fourier components of the scattered field) of the particle hit the poles of the reflection coefficient of the substrate, in which case the reflection becomes very strong. These poles correspond to the coupled surface plasmon states (also called slab modes) sustained at the dielectric-metal interfaces which can only be excited by the evanescent waves coming from the dipole particle. We know that the Fourier transform of the scattered fields from the dipole covers all the range of $k$ vectors of different directions. In order to hit as many poles as possible, the band (where the poles lie on) of the plasmon mode has to be very flat at some frequency. Such a flat band can be realized if the layer thickness $t_1$ and $t_2$ are properly chosen. Figure 2(a) shows the band diagram (TM mode) of the lossless ($\gamma = 0$) sandwich substrate with optimized parameters $t_1 = 0.1\lambda_p$, and $t_2 = 0.218\lambda_p$, where $\lambda_p = 2\pi c/\omega_p$, where a flat band is found at about $\omega = 0.25\omega_p$. At this frequency the Green's function is characterized by a resonance behavior as shown in Fig.2 (b) and (c), where we show the real and imaginary parts of $G_{xx}$ as a function of frequency $\omega$ and particle-substrate distance $d$. We note that the real part [Fig.2(b)] undergoes dramatic changes from very large positive value to very large negative value at the resonance frequency, indicating there is chance that $\left|1 - \mu_0 \omega^2 \alpha^e G_{xx}\right| \to 0$ and hence the effective polarizability is strongly enhanced. The imaginary part of $G_{xx}$ [Fig.2(c)] affects the quality factor of the resonance.

To calculate the optical force acting on the particle, the total electromagnetic field has to be determined first. After we obtained the dipole moment using Eq.(2), the field radiating from the dipole is then decomposed into different spectral components of plane wave form, which are then separately treated for their transmissions and reflections. The fields at the upper half space ($z > 0$) due to the particle can be expressed as [27]

$$E_z^p(\rho,\phi,z) = \frac{p_x \cos\phi}{4\pi\varepsilon_0} \int_0^\infty k_\rho^2 J_1(k_\rho \rho) \left[ \text{sign}(z-d) e^{ik_{0,z}|z-d|} - R^{\text{TM}} e^{ik_{0,z}(z+d)} \right] dk_\rho,$$

$$H_z^p(\rho,\phi,z) = \frac{\omega p_x \sin\phi}{4\pi} \int_0^\infty \frac{k_\rho^2}{k_{0,z}} J_1(k_\rho \rho) \left[ e^{ik_{0,z}|z-d|} + R^{\text{TE}} e^{ik_{0,z}(z+d)} \right] dk_\rho,$$

(3)

where $(\rho,\phi,z)$ denote the cylindrical coordinates; $J_1(x)$ is the cylindrical Bessel function; sign($x$) is a sign function defined as: $\text{sign}(x) = 1$ if $x \geq 0$, else $\text{sign}(x) = -1$. $k_\rho$ is the in-plane wavevector. $R^{\text{TE}}$ and $R^{\text{TM}}$ are the reflection coefficients of the sandwich substrate and their expressions can be found in Ref.[27]. After the longitudinal components are obtained using Eq.(3), the transverse field components can be determined with the following relations:

$$\tilde{\mathbf{E}}_t^p(k_\rho, \mathbf{r}) = \frac{1}{k_\rho^2}\left(\nabla_t \frac{\partial \tilde{E}_z^p}{\partial z} - i\mu\omega\hat{z} \times \nabla_t \tilde{H}_z^p\right),$$

$$\tilde{\mathbf{H}}_t^p(k_\rho, \mathbf{r}) = \frac{1}{k_\rho^2}\left(\nabla_t \frac{\partial \tilde{H}_z^p}{\partial z} + i\varepsilon\omega\hat{z} \times \nabla_t \tilde{E}_z^p\right).$$

(4)

Here $\nabla_t = \nabla - \hat{z}\partial/\partial z$ and $\tilde{E}_z^p, \tilde{H}_z^p$ denote the integrand of each spectral component in Eq. (3). Integrating the above equation over $k_\rho$ will give the transverse field components $\mathbf{E}_t^p, \mathbf{H}_t^p$. The time-averaged optical force exerted on the particle is then evaluated as:

$$\langle F_j \rangle = \frac{1}{2} \sum_i \mathrm{Re}\left[ \alpha_e E_i \partial_j E_i^* \right], \tag{5}$$

Where $E_i = E_i^p + E_i^{\mathrm{inc}}\left(1 + R^{\mathrm{TE}} e^{i2k_0 d}\right)$ is the total electric field acting on the particle, "*" denotes the complex conjugate and $\partial_j$ denotes the derivative with respect to $j$ Cartesian component. Equation (5) is used to determine the optical force acting on the particle in our model system, and for comparison we also calculated the force when the sandwich substrate is replaced by a semi-infinite metal substrate (with the same material properties). The results calculated by the above semi-analytical method are also compared with that given by a full-wave numerical method [28-30]. We found excellent agreement between our analytic theory and numerical results (shown in Sec.3).

### 3. Optical force enhancement due to particle-substrate resonance

*3.1. Particle with a negative dipole polarizability*

As the spherical particle is characterized by an electric polarizability $\alpha^e$, we will show that the force enhancement can be achieved for both cases of $\alpha^e < 0$ and $\alpha^e > 0$. For the former we consider a metal particle of radius $r = 15$ nm with relative permittivity $\varepsilon_p = -2$. The particle can have a negative polarizability in this case due to the resonant excitation of the dipole surface plasmon (the Fröhlich condition [12]). In order to show that the force enhancement is entirely due to the singular behavior associated with the flat-band feature but is not caused by the electric dipole resonance or some special material property of the particle, we fix $\varepsilon_p$ in the considered frequency region. For the metal layers of the substrate we set $\omega_p = 1.375 \times 10^{16}$ rad/s (corresponding to silver) and we consider first the lossless case with $\gamma = 0$. Figure 3(a) shows the optical force acting on the particle as a function of distance $d$ and frequency $\omega$, where a resonance is clearly noted (the red region). The resonance force is positive (repulsive), meaning that it can block the particle from approaching the substrate. This is due to the fact that the strong coupling field between the particle and the substrate results in a field gradient along –z direction and the particle has a negative polarizability ($\alpha^e < 0$), hence the gradient force is along +z direction. For comparison, Fig.3(b) shows the optical force acting on the same particle when the sandwich structure is replaced with a semi-infinite metal of the same material property. In this case, the near-field coupling is weak and the force is mainly induced by the standing wave due to the reflection of the incident plane wave. As the field gradient is along +z direction, the force is negative (attractive) in the considered frequency region as shown in Fig. 3(b). In order to show the "shielding" ability of the optical force, we then calculated the energy barrier which is defined as the integral of the optical force over distance $d \in [d_{min}, d_{max}]$, where $d_{min} = r$ is the minimum distance between the particle and the substrate surface while $d_{max}$ is truncated to 100 nm. The results in Fig.3(c) shows that the model system gives a positive energy barrier (at resonance) that is about 15 times larger than that of the semi-infinite metal case (refer to axis labels on the right-hand side) and this energy barrier can prevent the nano particle from getting close to the substrate. For the semi-infinite metal slab case, such a particle will be simply dragged to the substrate surface as the corresponding energy barrier has a negative value.

We now fix $d = 40$ nm and examine the detail of the force enhancement. Figure 4(a) shows the comparison between the optical force of the sandwich structure model system (denoted as $F$) and that of the semi-infinite metal system (denoted as $F_{\text{semi-infinite}}$). The axis labels on the left-hand side of Fig. 4(a) show the force normalized to the normal photon pressure $F_0$ acting on the same particle without the substrate. We see that the resonance force is about 3 orders of magnitude stronger than the normal photon pressure. In the same figure panel, the axis labels on the right hand side show the ratio of $F/|F_{\text{semi-infinite}}|$ and the maximum value is about 100. This demonstrates that the considered model system can give very strong enhancement of the coupling force between the particle and the substrate. To verify the results given by the above semi-analytical method, we also employ a full-wave method [28, 29] to calculate the force in the model system and the results are denoted as circles in Fig. 4(a), where the agreement is very good. Figure 4(b) shows the magnitude of the induced dipole moment $p = |\mathbf{p}|$ in the two cases, where about one order of magnitude of enhancement due to the resonance is observed. Figure 4 (c) and (d) shows the distributions of the resonant electric field amplitude on the *xoz*-plane and *yoz*-plane, respectively. We note the strong coupling field between the particle and the slab (the field values here correspond to the incident plane wave with amplitude $E_0 = 1$ V/m).

*3.2. Particle with a positive dipole polarizability*

We now keep the parameters of the sandwich substrate unchanged but consider the particle is made of dielectric material with $\varepsilon_p = 12.96$ so that $\alpha^e > 0$. We repeat the force calculation and the results are shown in Fig.5. Figure 5(a) shows the optical force acting on this dielectric particle as a function of distance $d$ and frequency $\omega$, where the resonance force now becomes attractive. This can also be explained by the same mechanism associated with the gradient force as for such a dielectric particle we have $\alpha^e > 0$ and the strong near-field due to coupling has a gradient along –z direction. Figure 5(b) shows the force in the semi-infinite metal system and we see it is repulsive in the considered frequency region because the field gradient produced by the standing wave is along +z direction. Figure 5(c) and (d) show the comparisons of the optical force and the induced dipole moment in the two cases. We see that enhancement of $F/|F_{\text{semi-infinite}}|$ can reach 50 at resonance and the induced dipole moment also is enhanced by one order of magnitude. We note that although the quality factor in this case is much larger than that of the $\alpha^e < 0$ case, its enhancement is relatively smaller. The reason is that the value of $|\alpha^e|$ for the dielectric case ($\varepsilon_p = 12.96$) is smaller than that of the metal case ($\varepsilon_p = -2$), therefore the residue of $|1 - \mu_0 \omega^2 \alpha^e G_{xx}|$ is larger and the effective polarizability is smaller.

*3.3. Effect of loss*

We have shown in Figs.3-5 that a dramatic force enhancement can be achieved for both $\alpha^e < 0$ and $\alpha^e > 0$ cases when the metal-dielectric-metal substrate is lossless. In general the material loss of the substrate could compromise such kind of resonance. Here we take the $\alpha^e < 0$ case as an example and study the effect of material loss when the metal layers of the slab is made of lossy silver with $\omega_p = 1.375 \times 10^{16}$ rad/s and $\gamma = 0.002\omega_p$ [31]. We re-did the calculation and the enhancements of both the optical force and the induced dipole moment are show in Fig. 6(a) and (b), respectively. Figure 6(a) shows that when loss is taken into account, the enhancement of the optical force (labels on the right-hand side) becomes smaller but it still can reach about one order of magnitude, compared to the case with non-structured semi-infinite metal substrate. The induced dipole moment is also several times larger [Fig. 6(b)].

## 4. Conclusions

In short, we have shown that in a simple particle-substrate configuration it is possible to induce very strong optical force acting on a nano-sized particle. The strong optical force is attributed to a particle-substrate resonance that is closely related to the divergence behavior of the effective dipole polarizability and the flat-band property of the metal-dielectric-metal slab modes. We show that by tuning the thickness of each layer of the sandwich-structured substrate one can obtain a very flat band, resulting in very high density of states at a certain frequency. Such a flat band can give very strong resonance coupling between the particle and the slab, and hence induce a significant optical force acting on a nano-sized particle which is orders of magnitude stronger than that of the non-structured system. We note that the flat-band property is also closely related to the origin of recently discovered super-scattering of light by sub-wavelength structures [32]. The physics here is not limited to spherical particles and is also applicable to non-spherical dipole particles. Our study may find applications in the optical force manipulations of very small particles on a substrate. The strong field enhancement inside the sandwich structure may be useful in light-harvest related applications. Besides, the enhancement of the dipole moment (which can make small particles much brighter) may also be applied in detecting small particles/molecules in similar systems.

**Acknowledgements**

This work is supported by Hong Kong RGC grant M-HKUST601/12. We thank Dr. M. Xiao for useful discussions.

*Correspondence to C. T. Chan (phchan@ust.hk)

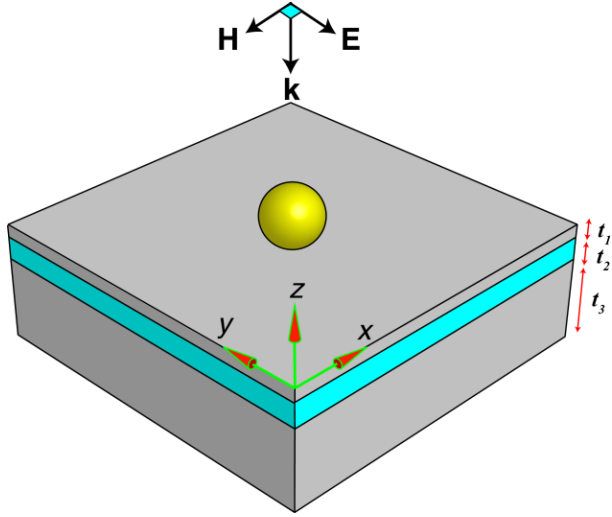

Fig.1. A schematic picture showing the configuration of the model system. A spherical particle of radius *r* is positioned at a distance of *d* above a sandwich-structured (metal-dielectric-metal) substrate. The layer thicknesses of the substrate are $t_1$, $t_2$, $t_3$.

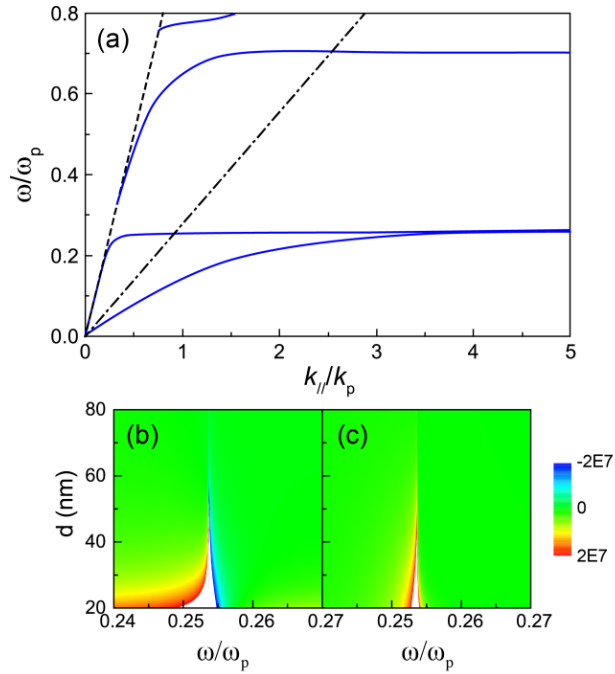

Fig.2. (a) Band diagram of the sandwich-structured substrate. The two dotted lines denote the light lines in vacuum and in the dielectric medium. Real [(b)] and imaginary [(c)] parts of the Green's function vs. frequency and distance *d*. The white region denotes off-scale values beyond that represented by the color bar.

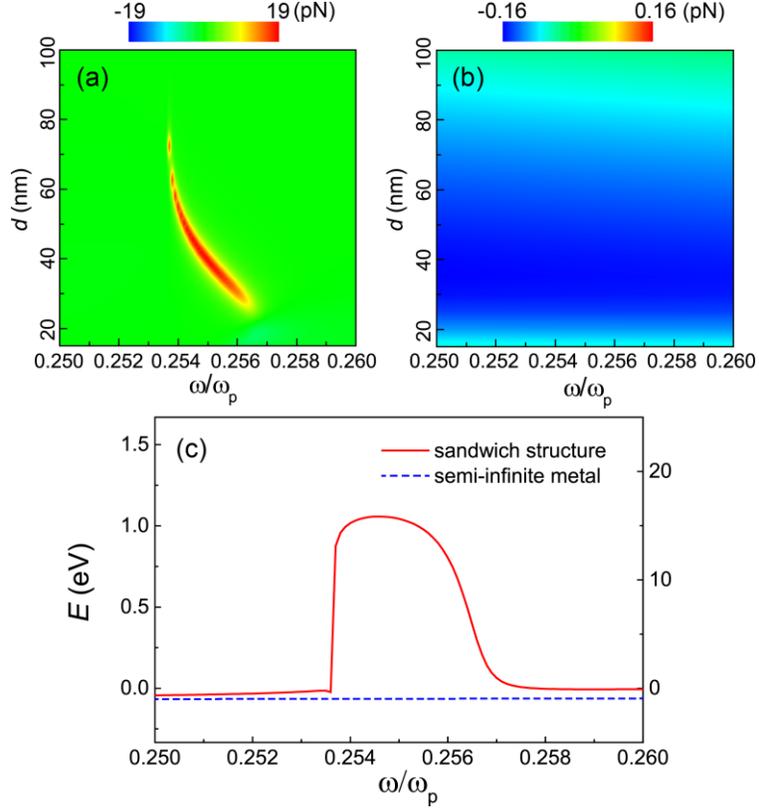

Fig.3. Optical force acting on a spherical particle with $\varepsilon_p = -2$ as a function of distance $d$ and frequency $\omega$ for the case of sandwich-structured substrate [(a)] and for the case of semi-infinite metal substrate [(b)]. (c) Comparison of the energy barrier. The axis labels on the left-hand side denote the energy, while those on the right-hand side denote the ratio between the two cases.

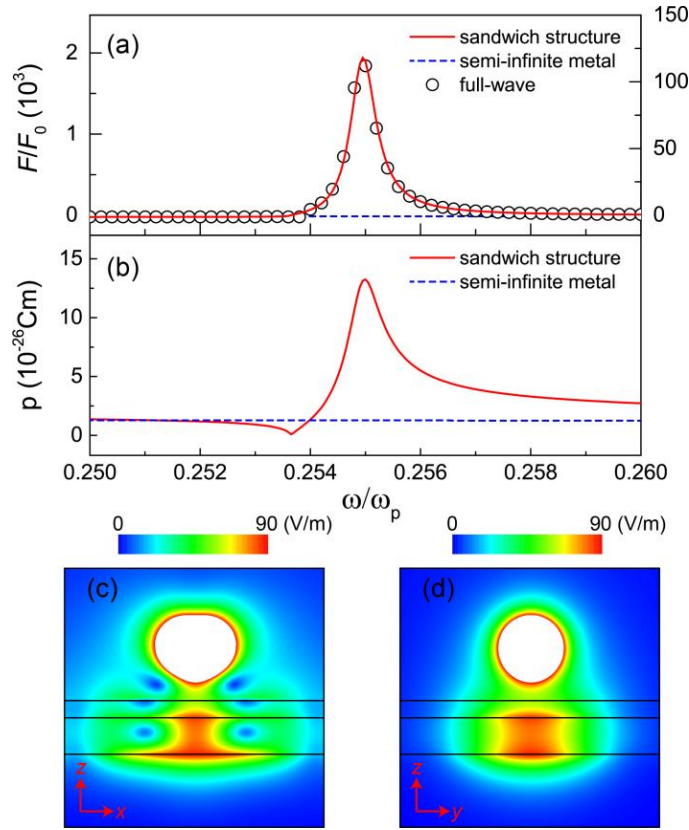

Fig.4. Comparison of the optical force [(a)] and the induced dipole moment [(b)] in the sandwich-structure substrate and the semi-infinite metal substrate. The circles in (a) denote the force computed using the full-wave method introduced in Ref.[29]. (c) Amplitude of the total electric field on the *xoz* plane. (d) Amplitude of the total electric field on the *yoz* plane.

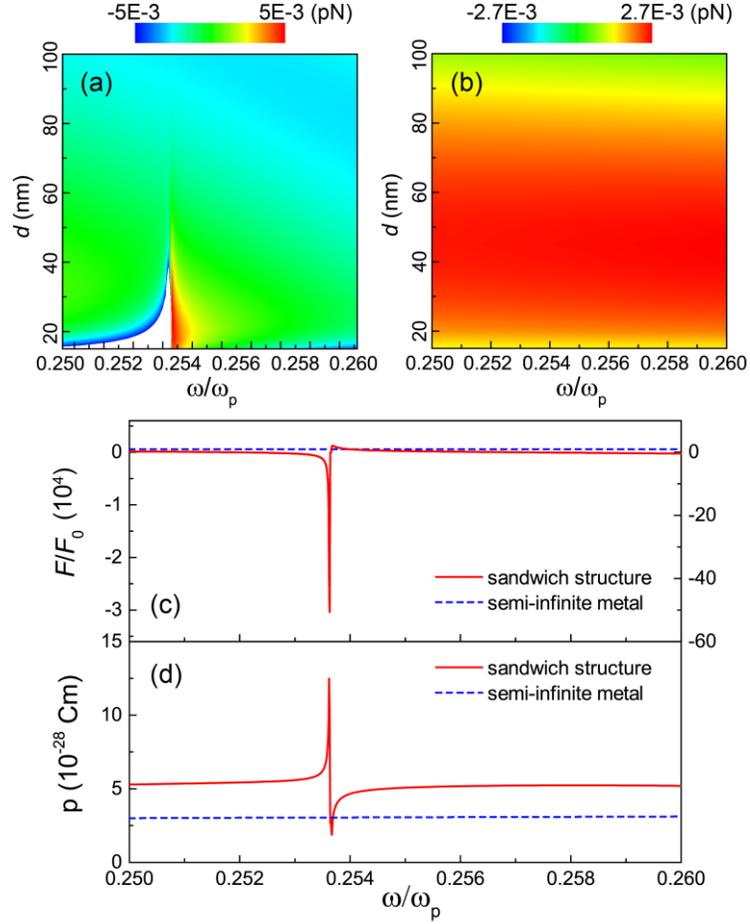

Fig.5. Optical force acting on a dielectric particle with $\varepsilon_p = 12.96$ as a function of distance $d$ and frequency for the sandwich structure case [(a)] and for the semi-infinite metallic substrate [(b)]. Comparisons of the force [(c)] and the induced dipole moment [(d)] between two considered cases.

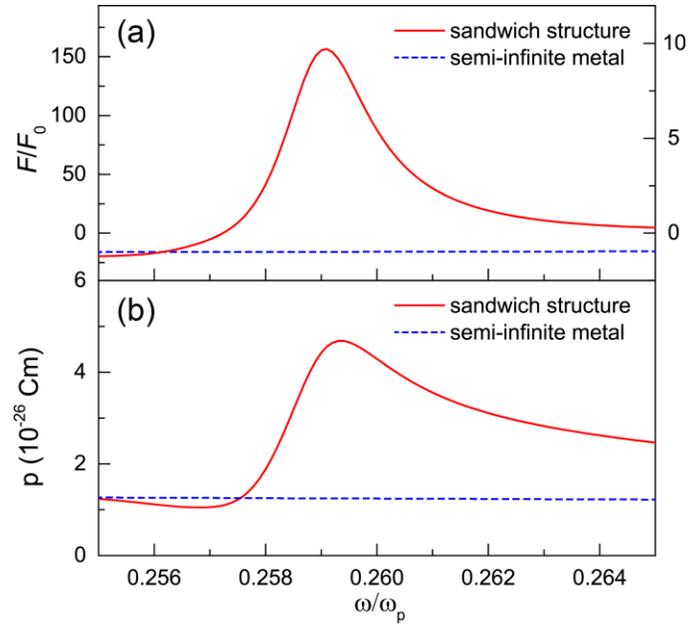

Fig. 6. The effect of loss on optical force enhancement. When the metal layers of the sandwich substrate is taken to be lossy silver, the enhancements of both the force (a) and the dipole moment (b) are reduced. But the resonant force is still one order of magnitude larger than that of the non-structured semi-infinite metal substrate case.